\documentclass[english,12pt]{article}
\usepackage[T1]{fontenc}
\usepackage[latin1]{inputenc}

\usepackage{geometry}
\usepackage{epsfig}
\usepackage{cite}
\usepackage{color}
\usepackage{comment}
\usepackage{color}
\usepackage{amsmath}

\newcommand\sss{\mathchoice%
{\displaystyle}%
{\scriptstyle}%
{\scriptscriptstyle}%
{\scriptscriptstyle}%
}

\def\({\left(} 
\def\){\right)} 
\newcommand\pt{p_{\sss\rm T}}

\begin{document}
\begin{titlepage}
{\begin{flushright}{
 \begin{minipage}{5cm}
   KA-TP-31-2008\\ SFB/CPP-08-95
\end{minipage}}\end{flushright}}
\vspace{1cm}
\begin{center} {\Large \bf VBFNLO: A parton level Monte Carlo
    } 
\end{center}
\begin{center}
 {\Large \bf  for processes with electroweak bosons}
\end{center}
\vspace{0.6cm}
\begin{center}
{\large \bf K. Arnold$^{1}$, M. B\"ahr$^{1}$, G. Bozzi$^{1}$, 
F. Campanario$^{1,2}$, C. Englert$^{1}$,}\\
\vspace{0.2cm}
{\large \bf T. Figy$^{3}$, N. Greiner$^{4}$, C. Hackstein$^{1}$, 
V. Hankele$^{1}$, B. J\"ager$^{5}$,}\\
 \vspace{0.2cm}
{\large \bf G. Kl\"amke$^{1}$, M. Kubocz$^{1}$, C. Oleari$^{6}$, 
S. Pl\"atzer$^{1}$, S. Prestel$^{1}$,}\\
\vspace{0.2cm}
{\large \bf  M. Worek$^{1,7}$, D. Zeppenfeld$^{1}$}
\end{center}
\vspace{0.4cm}
\begin{center}
$^{1}$ ITP, Universit\"at Karlsruhe, 76128 Karlsruhe,  Germany \\ \noindent
$^{2}$ Departament de F\'isica Te\'orica and IFIC, Universitat de
    Val\'encia - CSIC, \\ E-46100 Burjassot, Val\'encia, Spain \\ \noindent
$^{3}$ IPPP, University of Durham, Durham DH1 3LE, UK\\ \noindent
$^{4}$ Institut f\"ur Theoretische Physik, Universit\"at Z\"urich,
8057 Z\"urich, Switzerland\\ \noindent
$^{5}$ Institut f\"ur Theoretische Physik und Astrophysik, Universit\"at
W\"urzburg, 97074~W\"urzburg, Germany
\\ \noindent
$^{6}$ Universit\`a di Milano-Bicocca and INFN, Sezione di
    Milano-Bicocca,\\  20126 Milano, Italy \\ \noindent
$^{7}$ Institute of Physics, University of Silesia, 40-007 Katowice, Poland \\
\noindent
\end{center}
\vspace{0.6cm}
\begin{abstract}

\textsc{Vbfnlo} is a fully flexible parton level Monte Carlo program for the
simulation of vector boson fusion, double and triple vector boson
production in hadronic collisions at next-to-leading order in the strong
coupling constant. \textsc{Vbfnlo} includes Higgs and vector boson decays
with full spin correlations and all off-shell effects.  In addition,
\textsc{Vbfnlo} implements ${\cal{CP}}$-even and ${\cal{CP}}$-odd Higgs boson
via gluon fusion, associated with two jets, at the leading-order one-loop level
with the full top- and bottom-quark mass dependence in a generic
two-Higgs-doublet model.

A variety of effects arising from beyond the Standard Model physics are
implemented for selected processes. This includes anomalous couplings of
Higgs and vector bosons and a Warped Higgsless extra dimension model.  The
program offers the possibility to generate Les Houches Accord event files 
for all processes available at leading order.
\end{abstract}
\vspace{2 cm}
\today
\end{titlepage}

\newpage

{\bf Program summary}

\vspace{0.5cm}

\begin{small}
\noindent
{\em Program Title:}  \textsc{Vbfnlo}    
                                    \\
{\em Journal Reference:}                                      \\
{\em Catalogue identifier:}                                   \\
{\em Licensing provisions:}  GPL version 2                                \\
{\em Program obtainable from:} 
{\tt http://www-itp.particle.uni-karlsruhe.de/vbfnlo/}        \\
{\em Distributed format:}  tar gzip file                                  \\
{\em Programming language:} Fortran, parts in C++                   \\
{\em Computer:} All                                              \\
{\em Operating system:}  Linux, should also work on other systems        
{\em Classification:} 11.1, 11.2  \\  
{\em External routines/libraries:}  
Optionally Les Houches Accord PDF Interface library 
and the GNU Scientific library. \\
 {\em Nature of problem:} 
 To resolve the large scale dependence inherent in
 leading order calculations and to quantify the cross section error
 induced by uncertainties in the determination of parton distribution 
 functions, it is necessary to include NLO
 corrections. Moreover, whenever  stringent cuts are required on decay 
 products and/or identified jets the question arises whether the scale 
 dependence and a k-factor,
 defined as the ratio of NLO to LO cross section, determined for the inclusive
 production cross sections are valid for the search region one is interested 
 in. \\ 
{\em Solution method:} The problem is best addressed by implementing the
one-loop QCD corrections in a fully flexible NLO parton-level Monte Carlo 
program, where arbitrary cuts can be specified as well 
as various scale choices. In addition, any currently available 
parton distribution 
function set can be used through the LHAPDF library.\\
{\em Running time:} Depending on the process studied. Usually from minutes
to hours. 
\end{small}

\tableofcontents


\section{Introduction}

The physics potential of the \textsc{TeVatron} and even more of
the \textsc{Lhc} relies, to a large extent,
on our ability to provide accurate cross
section predictions both for signal and background processes. The latter 
are often generated by QCD interactions followed  by weak transitions
of partons to vector bosons. A precise 
description of such hard QCD production processes 
is needed, as well as a method 
for simulating the measurable hadronic final states. 
Reaching these goals requires next-to-leading order (NLO) QCD
calculations presented in the form of parton level Monte Carlo~(MC) generators
 which are  an efficient solution when it comes to final states
characterized by a high number of jets and/or identified particles.
When kinematical cuts are imposed, as is mandatory for processes
involving QCD radiation, analytical phase space integration becomes
impractical and implementation of results in the form of Monte Carlo
programs becomes the method of choice. 

\textsc{Vbfnlo} is  a fully flexible MC program for 
vector boson fusion (VBF), double and triple vector boson 
production processes at NLO QCD accuracy.  Since real emission processes
are part of the NLO cross sections, \textsc{Vbfnlo} provides the
possibility to calculate cross sections for the corresponding process 
with  one additional jet at leading order (LO) in the strong QCD coupling.
In addition, the simulation of $\cal{CP}$-even and $\cal{CP}$-odd Higgs
boson production in gluon fusion, 
associated with two additional jets, is implemented at leading order 
in the strong coupling
with the full top- and bottom-quark mass dependence in a generic
two-Higgs-doublet model. Several models for anomalous couplings of
Higgs- and vector bosons and a Warped Higgsless 
extra dimension model have been implemented.

Arbitrary cuts can be specified as well as various scale choices. Any
currently available parton distribution function (PDF) set can be used
through the LHAPDF library\footnote{\tt http://projects.hepforge.org/lhapdf/}. 
For processes implemented at leading order, the
program is capable of generating event files in the Les Houches Accord (LHA)
format \cite{Alwall:2006yp}.


\section{Processes}
\label{sec:proc}
In the following sections, we describe all production processes and decay
modes implemented in \textsc{Vbfnlo}, together with references to a more
detailed discussion of the underlying calculations.

In the phase space regions which are accessible at hadron colliders, VBF
reactions are dominated by $t$-channel electroweak gauge boson exchange. In
\textsc{Vbfnlo} therefore $s$-channel exchange contributions and kinematically
suppressed fermion interference contributions
\cite{CO,Andersen:2007mp,Bredenstein:2008tm} are disregarded. 
Throughout, we consider only decays of the weak bosons into different lepton
generations, such as $ZZ\to \ell^+\ell^-\ell'^+\ell'^-$. Results for leptonic
final states with any combination of leptons (e.g.,
$\ell^+\ell^-\ell^+\ell^-$) can be obtained thereof by multiplying the
respective results with appropriate combinatorial factors. Numerically small
contributions from Pauli-interference effects for identical charged leptons
are disregarded.  

\subsection{VBF Higgs production in association with two jets}
\label{sec:vbf-hjj}

$Hjj$ production via VBF mainly proceeds via electroweak quark-quark
scattering processes like $qq'\to qq'H$ and crossing related reactions. 
In \textsc{Vbfnlo}, tree level cross sections and NLO QCD corrections to 
the $t$-channel production process are provided. 
The subsequent decay of the Higgs boson 
is simulated in the narrow width approximation (NWA). 
For the $H\to W^+W^- $ and the $H\to ZZ$ modes, full off-shell
effects and spin correlations of the decay leptons are included.   
The available production process and the decay modes are listed with the
corresponding process IDs in Table~\ref{tab:prc1}.   
Details of the calculation can be found in Ref.~\cite{Figy:2003nv}.
\begin{table}[h!]
\newcommand{\lstrut}{{$\strut\atop\strut$}}
\begin{center}
\small
\begin{tabular}{c|l}
\hline
&\\
\textsc{ProcId} & \textsc{Process}  \\
&\\
\hline
&\\
\bf 100 & $pp\to H \, jj$  \\
\bf 101 & $pp\to H \, jj\to \gamma\gamma \, jj$ \\
\bf 102 & $pp\to H \, jj\to \mu^+\mu^- \, jj$ \\
\bf 103 & $pp\to H \, jj\to \tau^+\tau^- \, jj$ \\
\bf 104 & $pp\to H \, jj\to b\bar{b} \, jj$ \\
\bf 105 & $pp\to H \, jj\to W^+W^- \, jj\to \ell^+\nu_\ell \ell'^- 
\bar{\nu}_{\ell'} \,jj$ \\
\bf 106 & $pp\to H \, jj\to ZZ \, jj\to \ell^+ \ell^- \ell'^+ 
\ell'^- \,jj$ \\
\bf 107 & $pp\to H \, jj\to ZZ \, jj\to \ell^+ \ell^- \nu_{\ell'}  
\bar{\nu}_{\ell'} \,jj$ \\
&\\
\hline
\end{tabular}
\caption {\em  Process IDs for $pp\to Hjj$ production via
  weak boson fusion at NLO QCD accuracy. }
\vspace{0.2cm}
\label{tab:prc1}
\end{center}
\end{table}

\subsection{VBF Higgs production in association with three jets}

Adding an extra parton to the Higgs production processes of
Sec.~\ref{sec:vbf-hjj} gives
rise to $Hjjj$ final states. The corresponding cross sections are implemented
at NLO QCD accuracy in \textsc{Vbfnlo}.  A list of all available modes and
corresponding process IDs is given in Table~\ref{tab:prc2}.  Details of the
calculation can be found in Ref.~\cite{Figy:2007kv}.
\begin{table}[t!]
\newcommand{\lstrut}{{$\strut\atop\strut$}}
\begin{center}
\small
\begin{tabular}{c|l}
\hline
&\\
\textsc{ProcId} & \textsc{Process}  \\
&\\
\hline
&\\
\bf 110 & $pp\to H \, jjj$ \\
\bf 111 & $pp\to H \, jjj\to \gamma\gamma \, jjj$ \\
\bf 112 & $pp\to H \, jjj\to \mu^+\mu^- \, jjj$ \\
\bf 113 & $pp\to H \, jjj\to \tau^+\tau^- \, jjj$ \\
\bf 114 & $pp\to H \, jjj\to b\bar{b} \, jjj$ \\
\bf 115 & $pp\to H \, jjj\to W^+W^- \, jjj\to \ell^+\nu_\ell \ell'^- 
\bar{\nu}_{\ell'} \,jjj$ \\
\bf 116 & $pp\to H \, jjj\to ZZ \, jjj\to \ell^+ \ell^- \ell'^+ \ell'^- \,jjj$ \\
\bf 117 & $pp\to H \, jjj\to ZZ \, jjj\to \ell^+ \ell^- \nu_{\ell'}  
\bar{\nu}_{\ell'} \,jjj$ \\
&\\
\hline
\end{tabular}
\caption {\em  Process IDs for $pp\to Hjjj$ production via weak
  boson fusion at NLO QCD accuracy.}
\vspace{0.2cm}
\label{tab:prc2}
\end{center}
\end{table}

\subsection{VBF production of a vector boson and two jets}

Vector boson fusion processes can  also produce final states with two leptons
plus two jets, which are generically referred to as ``VBF $Zjj$
and $W^\pm jj$ production''. These reactions are implemented to NLO QCD
accuracy in \textsc{Vbfnlo}, see Table~\ref{tab:prc3}.  Details of the
calculation can be found in Ref.~\cite{Oleari:2003tc}.
\begin{table}[t!]
\newcommand{\lstrut}{{$\strut\atop\strut$}}
\begin{center}
\small
\begin{tabular}{c|l}
\hline
&\\
\textsc{ProcId} & \textsc{Process} \\
&\\
\hline
&\\
\bf 120 & $pp\to Z \, jj \to \ell^{+} \ell^{-} \, jj$ \\
\bf 121 & $pp\to Z  \, jj\to \nu_\ell \bar{\nu}_\ell \, jj$ \\
\bf 130 & $pp\to W^{+} \,  jj\to \ell^{+} \nu_\ell \, jj$ \\
\bf 140 & $pp\to W^{-} \, jj\to \ell^{-} \bar{\nu}_\ell  \, jj$ \\
&\\
\hline
\end{tabular}
\caption {\em  Process IDs for $Zjj$ and $W^{\pm}jj$  production 
via weak  boson fusion at NLO QCD accuracy.}
\vspace{0.2cm}
\label{tab:prc3}
\end{center}
\end{table}

\subsection{VBF production of two vector bosons and two jets}

The production of four leptons plus two jets in the final states at order
$\mathcal{O}(\alpha^6)$ is dominated by VBF contributions. In
\textsc{Vbfnlo}, all resonant and non-resonant $t$-channel exchange
contributions giving rise to a specific leptonic final state are
considered. For simplicity, we refer to these reactions as ``VBF $W^+W^-jj$,
$ZZjj$, and $W^\pm Zjj$ production''. Finite width effects of the weak bosons
and spin correlations of the decay leptons are fully retained.
The available processes and corresponding process  IDs are listed in  
Table~\ref{tab:prc4}. Details of the calculation can be found in 
Refs.~\cite{Jager:2006zc,Jager:2006cp,Bozzi:2007ur}.
\begin{table}[t!]
\newcommand{\lstrut}{{$\strut\atop\strut$}}
\begin{center}
\small
\begin{tabular}{c|l}
\hline
&\\
\textsc{ProcId} & \textsc{Process} \\
&\\
\hline
&\\
\bf 200 & $pp\to W^{+}W^{-} \, jj \to \ell^{+} \nu_\ell \ell'^{-}
\bar{\nu}_{\ell'} \, jj$ \\
\bf 210 & $pp\to ZZ  \, jj\to \ell^{+} \ell^{-} \ell'^{+} \ell'^{-} \, jj$ \\
\bf 211 & $pp\to ZZ  \, jj\to \ell^{+} \ell^{-} \nu_{\ell'} \bar{\nu}_{\ell'} \, jj$ \\
\bf 220 & $pp\to W^{+}Z \,  jj\to \ell^{+} \nu_\ell \ell'^{+} \ell'^{-} \, jj$ \\
\bf 230 & $pp\to W^{-}Z \, jj\to \ell^{-} \bar{\nu}_\ell 
\ell'^{+} \ell'^{-} \, jj$ \\
&\\
\hline
\end{tabular}
\caption {\em  Process IDs for  $W^{+}W^{-}jj$, $ZZjj$ and $W^{\pm}Zjj$ 
production via weak  boson fusion at NLO QCD accuracy.}
\vspace{0.2cm}
\label{tab:prc4}
\end{center}
\end{table}

\subsection{Double and triple vector boson production}

The production of four- and six-lepton final states mainly proceeds via double and
triple vector boson production with subsequent decays. In \textsc{Vbfnlo}, the
processes listed in Table~\ref{tab:prc5} are implemented to NLO QCD accuracy,
including full off-shell effects and spin correlations of the final state
leptons.  Details of the calculation can be found in
Refs.~\cite{Hankele:2007sb,Campanario:2008yg}.
\begin{table}[t!]
\newcommand{\lstrut}{{$\strut\atop\strut$}}
\begin{center}
\small
\begin{tabular}{c|l}
\hline
&\\
\textsc{ProcId} & \textsc{Process}  \\
&\\
\hline
&\\
\bf 300 & $pp\to W^{+}W^{-} \to \ell_1^{+} \nu_{\ell_1} \ell_2^{-}\bar{\nu}_{\ell_2} $ \\
\bf 400 & $pp\to W^{+}W^{-}Z \to {\ell_1}^{+}\nu_{\ell_1} {\ell_2}^{-} \bar{\nu}_{\ell_2} 
{\ell_3}^{+} {\ell_3}^{-} $ \\
\bf 410 & $pp\to ZZW^{+} \to  {\ell_1}^{+} {\ell_1}^{-}  {\ell_2}^{+} {\ell_2}^{-} 
 {\ell_3}^{+} \nu_{\ell_3} $ \\
\bf 420 & $pp\to ZZW^{-} \to {\ell_1}^{+} {\ell_1}^{-}  {\ell_2}^{+} {\ell_2}^{-} 
 {\ell_3}^{-}  \bar{\nu}_{\ell_3}$ \\
\bf 430 & $pp\to W^{+}W^{-}W^{+} \to {\ell_1}^{+}\nu_{\ell_1} {\ell_2}^{-}
\bar{\nu}_{\ell_2} {\ell_3}^{+}\nu_{\ell_3}$ \\
\bf 440 & $pp\to W^{-}W^{+}W^{-} \to {\ell_1}^{-} \bar{\nu}_{\ell_1}{\ell_2}^{+}\nu_{\ell_2}
{\ell_3}^{-} \bar{\nu}_{\ell_3} $ \\
&\\
\hline
\end{tabular}
\caption {\em  Process IDs for the $W^{+}W^{-}$, $WWZ$, $ZZW$ and $WWW$
  production processes at NLO QCD accuracy.}
\vspace{0.2cm}
\label{tab:prc5}
\end{center}
\end{table}
%

\subsection{Higgs production in gluon fusion with two jets}

%
$\cal{CP}$-even and $\cal{CP}$-odd Higgs boson production in gluon fusion,
associated with two additional jets, is a process which first appears at the
1-loop level which, therefore, is counted as leading order in the strong
coupling.  This process is simulated including the full mass dependence of
the top and bottom quark running in the loop of a generic two-Higgs-doublet
model.
\begin{table}[t!]
\newcommand{\lstrut}{{$\strut\atop\strut$}}
\begin{center}
\small
\begin{tabular}{c|l}
\hline
&\\
\textsc{ProcId} & \textsc{Process} \\
&\\
\hline
&\\
\bf 4100 & $pp\to H \, jj $ \\
&\\
\hline
\end{tabular}
\caption {\em  Process ID for the LO Higgs plus 2 jets
  production via gluon fusion.}
\vspace{0.2cm}
\label{tab:prc6}
\end{center}
\end{table}
The relevant process ID  is given in Table~\ref{tab:prc6}. 
Details of the calculation can be found in
 Refs.~\cite{DelDuca:2001eu,DelDuca:2001fn,DelDuca:2006hk,Klamke:2007cu,GF}.


\section{Installing VBFNLO}

The source code of the current version of \textsc{Vbfnlo} can be
downloaded from the \textsc{Vbfnlo} web-page
\begin{center}
{\tt  http://www-itp.particle.uni-karlsruhe.de/vbfnlo/}
\end{center}
and includes a GNU conforming build system for portability and an
easy build and installation procedure.

\subsection{Prerequisites}

The basic installation requires GNU {\tt make}, a 
\textsc{Fortran77}\footnote{The
  following compilers have been tested: {\tt g77} and {\tt gfortran}} 
 and a C++ compiler. \textsc{Vbfnlo} offers the possibility to use 
the LHAPDF\footnote{\tt http://projects.hepforge.org/lhapdf/} 
library for parton distribution functions. In case the simulation of
Kaluza-Klein resonances should be enabled, an installation of the GNU
Scientific Library (GSL)\footnote{\tt http://www.gnu.org/software/gsl/}
is required.


\subsection{Build and installation}

After unpacking the source archive and entering the source directory, the {\tt
configure} script can be invoked with several options, a complete list being
available via {\tt ./configure --help}.  Among these, the most important ones
are:

\begin{itemize}
\item {\tt ----prefix=[path]} \\ Install \textsc{Vbfnlo} in the location given by {\tt
[path]}.
\item {\tt ----enable-processes=[list]} \\ By default, the code for all
  available 
processes is compiled. Optionally, {\tt [list]} gives a comma-separated list of
selected processes to be compiled. Possible process names are:

\begin{tabular}{ll} {\tt vbf} & Vector boson fusion processes \\ {\tt diboson}
& Double gauge boson production \\ {\tt triboson} & Triple gauge boson
production \\ {\tt hjjj} & Higgs boson plus three jet production in vector
boson 
fusion \\ {\tt ggf} & Higgs boson plus two jets via gluon fusion
\end{tabular}

\item {\tt ----disable-NLO} \\ Disable the next-to-leading order QCD
corrections.  With this option, compilation time is shortened.
\item {\tt ----enable-kk} \\ Enable simulation of Kaluza-Klein resonances.
Disabled by default, the Kaluza-Klein option requires the installation of the GNU Scientific
Library to be specified via {\tt ----with-gsl}.
\item {\tt ----with-LHAPDF=[path]} \\ Enable the usage of LHAPDF instead of the
built-in PDF sets.  Disabled by default. {\tt [path]} specifies the location of
the LHAPDF installation.
\item {\tt ----with-gsl=[path]} \\ Enable usage of the GNU Scientific
Library. {\tt [path]} specifies the location of the GSL installation.
\end{itemize}

Once {\tt configure} finished successfully, the {\tt make} and {\tt make
install} commands will compile and install \textsc{Vbfnlo}, respectively.


\subsection{Special notes for Mac OS X}

Due to a linker problem on this platform, \textsc{Vbfnlo} has to be compiled 
with
static libraries disabled by adding the \texttt{----disable-static} flag to
the call of the configure script. For the dynamic libraries to be properly
resolved, the environment variable \texttt{DYLD\_LIBRARY\_PATH} has to be
set to the library directory of the installation:
\begin{verbatim}
export DYLD_LIBRARY_PATH=[prefix]/lib/VBFNLO
\end{verbatim} 
where
{\tt [prefix]} is the installation directory as chosen by the
{\tt ----prefix} parameter.


\subsection{Source and installation directory layout}

The \textsc{Vbfnlo} source tree contains the following subdirectories:
\begin{itemize}
\item {\tt amplitudes/:} Routines to calculate matrix elements for the 
processes provided.
\item {\tt doc/:} The source of this manual.
\item {\tt helas/:} \textsc{Helas} \cite{Murayama:1992gi} subroutines used
to calculate helicity amplitudes.
\item {\tt loops/:} One-loop tensor integrals up to five-point functions.
\item {\tt PDFsets/:} Built-in parton distributions (CTEQ6L1 and 
CTEQ6M, \cite{Pumplin:2002vw}).
\item {\tt phasespace/:} Specialized phasespace generators for the processes provided.
\item {\tt src/:} Source code of the main programs and input files. 
\item {\tt utilities/:} Routines for administrative tasks, cuts, scale choices and interfaces.
\end{itemize} 

The source does not need to be modified to change the simulation parameters.
\textsc{Vbfnlo} offers several kinematical cuts and scale choices. This is
illustrated in Sec.~\ref{sec:input}. In addition, it provides a few basic
histograms. Cuts, histograms and scale choices not already provided may be
added in the {\tt utilities/cuts.F}, {\tt utilities/histograms.F} and {\tt
utilities/scales.F} files.

The installation is performed in a standard \textsc{Unix}-layout, i.e.\ the directory
specified with the {\tt ----prefix} option of the {\tt configure} script contains the
following directories:
\begin{itemize}
\item {\tt bin/:} {\tt vbfnlo} and {\tt ggflo} executables.
\item {\tt include/VBFNLO/:} \textsc{Vbfnlo} header files. 
\item {\tt lib/VBFNLO/:} \textsc{Vbfnlo} modules as dynamically loadable
  libraries. These can also
be used independently from one of the main programs.
\item {\tt share/VBFNLO/:} Input files and internal PDF tables.
\end{itemize}


\subsection{Running the program}

Both the {\tt vbfnlo} and {\tt ggflo} executables contained in the
{\tt bin} directory of the installation path do look for input files
in their current working directory. An alternative path to input files
may be specified explicitly by passing the {\tt input=path} argument
to the programs, with {\tt path} denoting the full path where input files
are located.

The input files contained in the {\tt share/VBFNLO} directory are meant to represent
default settings and should not be changed.
We therefore recommend to symbolically link the desired executable and copy the
input files to a separate directory. Here, special settings may be chosen in the
input files and the program can be run in that directory without specifying further
options.


\subsection{Bug reports}

Please report any problems to
\begin{verbatim}
  vbfnlo@particle.uni-karlsruhe.de
\end{verbatim}
including a short report with which configure options \textsc{Vbfnlo} has been built,
as well as the versions of compilers and external libraries used.

\subsection{License}

\textsc{Vbfnlo} is distributed under the GNU General Public License (GPL)
version~2. This ensures that the source code will be available to users,
grants them the freedom to use and modify the program and sets
out the conditions under which it can be redistributed. However, it was
developed as part of an academic research project and is the result of
many years of work by the authors, which raises various issues that are
not covered by the legal framework of the GPL. It is therefore
distributed together with a set of guidelines\footnote{These guidelines
are contained in the \texttt{GUIDELINES} file distributed with the
release.}, which originally have been formulated and agreed on by the
MCnet collaboration for event generator software.


\section{Input files and parameters}
\label{sec:input} 
\textsc{Vbfnlo} is steered through the following input files:
\begin{itemize}
\item {\tt vbfnlo.dat:} General parameters for a run.
\item {\tt ggflo.dat:} Additional parameters for the {\tt ggflo} program.
\item {\tt cuts.dat:} Values for kinematical cuts.
\item {\tt anom\_HVV.dat:} Parameters for anomalous Higgs couplings.
\item {\tt anom\_WW.dat:} Parameters for anomalous triple gauge boson
couplings.
\item {\tt random.dat:} Seed for the random number generator.
\end{itemize}

The following subsections will give a detailed description of all available parameters.


\subsection{{\tt vbfnlo.dat} $-$ general parameters}
\label{sec:general}
\begin{itemize}
\item {\tt PROCESS:} Process ID as described in Sec.~\ref{sec:proc}.

\item {\tt LOPROCESS\_PLUS\_JET:} If set to true, the leading order
  process with one additional jet is generated, i.e. only the real radiation
  contribution is generated.  This option is available for all but gluon
  fusion processes. 
\item{\tt LEPTONS:} Choice of the final state leptons according to the MC
particle numbering scheme \cite{Amsler:2008zz}. If the selected
configuration is not available, default values are used.
\item {\tt LO\_ITERATIONS:} Sets the number of iterations for the integration
of LO cross sections. Usually more than one iteration is used in
order to adapt the integration grid and thus improve the efficiency of the
MC integration algorithm\footnote{For all NLO calculations the virtual
  contributions are calculated using the already optimized leading order grid.}. 
For an adapted grid file (see {\tt LO\_GRID})
this parameter can be set to 1. Default is 4.
\item {\tt NLO\_ITERATIONS:} Analogous to  {\tt LO\_ITERATIONS}, but for the real
emission part of an NLO calculation. Since the corresponding phase space 
is different from the LO configuration, a second independent
MC integration has to be performed. Default is 4.
\item {\tt LO\_POINTS:} Determines the number of phase space points that are
generated in each iteration. In the last iteration there are $2^N$ points,
where $N$={\tt LO\_POINTS}. In each previous iteration, the number of points
is half the value of the following one. Example: For 4 iterations ({\tt
LO\_ITERATIONS = 4}) and {\tt LO\_POINTS = 20}, there are $2^{17}$ generated
points in the first, $2^{18}$ in the second , $2^{19}$ in the third and
$2^{20}\approx 10^6$ in the last iteration\footnote{The virtual contributions
  are calculated for $2^{N}$ points only.}. 
Default is $N={20}$.
\item {\tt NLO\_POINTS:} Similar to {\tt LO\_POINTS}, but for the real emission part of
a NLO calculation.
\item {\tt LO\_GRID:} Sets the name of the grid files that are generated at
the end of each iteration. Choosing {\tt name} as input parameter, in each 
iteration $X$ a grid file {\tt name.out.X} will be produced. If a grid file 
{\tt name} is already present in the working directory, the program reads in 
this file when executed.
\item {\tt NLO\_GRID:} Similar to {\tt LO\_GRID}, but for the real emission part of a
NLO calculation.
\item {\tt NLO\_SWITCH:} Switch for the NLO part of a process, if available.
If set to {\tt .true.}, cross sections and histograms are calculated to NLO
QCD  accuracy. Default is set to {\tt .false.}.
\item {\tt ECM:} The center-of-mass energy $\sqrt{s}$ of the collider,
measured in GeV. Default is 14000 GeV.
\item {\tt BEAM1, BEAM2:} Define the type of particle of each beam. Possible
options are $+1$ for proton beams and $-1$ for anti-proton beams. Default
is proton-proton collisions, ($+1$, $+1$).
\item {\tt ID\_MUF:} Choice of the factorization scale. 
See Table~\ref{tab:fscales} for a list of available options. Default is 0.
\item {\tt ID\_MUR:} Choice of the renormalization scale. 
See Table~\ref{tab:rscales} for a list of available options. Default is 0.
\begin{table}[hhh]
\newcommand{\lstrut}{{$\strut\atop\strut$}}
\begin{center}
\begin{tabular}{c|p{10cm}|c}
\hline
&&\\
{\tt ID\_MUF } & \textsc{Factorization Scale} & \textsc{Process class} \\
&&\\
\hline
&&\\
\bf 0 & user defined constant scale set by {\tt MUF\_USER} & {\tt all}\\
\bf 1 & momentum transfer of exchanged $W/Z$ boson &  {\tt vbf} \\
\bf 2 & $\min(\pt(j_{1}),\pt(j_{2})$) & {\tt vbf} \\
\bf 3 & invariant mass of $VV$ system & {\tt diboson} \\ 
\bf 4 & invariant mass of $VVV$ system & {\tt triboson} \\ 
\bf 5 &  $\sqrt{ \pt(j_1) \times \pt(j_2)}$ & {\tt ggf}\\ 
&&\\
\hline
\end{tabular}
\caption {Factorization scale options. }
\vspace{0.2cm}
\label{tab:fscales}
\end{center}
\end{table}
\begin{table}[hhh]
\newcommand{\lstrut}{{$\strut\atop\strut$}}
\begin{center}
\begin{tabular}{c|p{10cm}|c}
\hline
&&\\
{\tt ID\_MUR } & \textsc{Renormalization Scale} & \textsc{Process class} \\
&&\\
\hline
&&\\
\bf 0 & user defined constant scale set by {\tt MUR\_USER} & {\tt all}\\
\bf 1 & momentum transfer of exchanged $W/Z$ boson &  {\tt vbf} \\
\bf 2 & $\min(\pt(j_{1}),\pt(j_{2})$) & {\tt vbf} \\
\bf 3 & invariant mass of $VV$ system & {\tt diboson} \\ 
\bf 4 & invariant mass of $VVV$ system & {\tt triboson} \\ 
\bf 5 &  $\alpha^{4}_s=\alpha_s(\pt(j_{1})) 
\times \alpha_s(\pt(j_{2})) \times \alpha^{2}_s(m_{H}) $ & {\tt ggf}\\ 
&&\\
\hline
\end{tabular}
\caption {Renormalization scale options. }
\vspace{0.2cm}
\label{tab:rscales}
\end{center}
\end{table}
\item {\tt MUF\_USER:} If {\tt ID\_MUF} is set to 0, this parameter sets the
user defined constant factorization scale measured in GeV. Default is 100 GeV.
\item {\tt MUR\_USER:} If {\tt ID\_MUR} is set to 0, this parameter sets the
user defined constant renormalization scale measured in GeV. Default is 100
GeV.
\item {\tt XIF:} Factor by which the
factorization scale is multiplied. May be used
to analyze the scale dependence of differential cross sections. Default is $1.0$.
\item {\tt XIR:} Factor by which the
renormalization scale is multiplied. May be used
to analyze the scale dependence of differential cross sections. Default is $1.0$.
\end{itemize}


\subsection{{\tt vbfnlo.dat} $-$ physics parameters}
\label{sec:physics}

\begin{itemize}
\item {\tt HMASS:} Standard Model Higgs boson mass in GeV. Default value is 120 GeV. 
\item {\tt TOPMASS:} Top quark mass in GeV. Default value is 172.4 GeV.
\item {\tt BOTTOMMASS:} Bottom quark pole mass in GeV, used in the calculation of
  the Higgs width and branching ratios as well as in the heavy quark loop 
in the gluon fusion process. Default value is 4.855 GeV.
\item {\tt CHARMMASS:} Charm quark pole mass in GeV used in the calculation of
  the Higgs width and branching ratios. Default value is 1.65 GeV.
\item {\tt ALFA\_S:} Strong coupling constant as used in the calculation of W, Z
  and H widths. The strong coupling constant used in the matrix element
  calculations is printed out during run time. Default value is 0.1176.
\item {\tt EWSCHEME:} Sets the scheme for the calculation of 
electroweak parameters. A summary of the four available options is given in Table~\ref{tab:Schemes}. 
Default value is 3.
\begin{table}[t!]
\newcommand{\lstrut}{{$\strut\atop\strut$}}
\begin{center}
\begin{tabular}{c|c|l|c}
\hline
&&&\\
{\tt EWSCHEME} & \textsc{Parameter} & \textsc{Default Value} & \textsc{Input/Calculated} \\
&&&\\
\hline
&&&\\
 & {\tt FERMI\_CONST} & $1.16637\times 10^{-5} \ \mathrm{GeV}^{-2}$ & \textsc{Input}\\
 & {\tt ALFA} & $ 7.2973525376\times 10^{-3}$ & \textsc{Input}\\
\bf 1 & {\tt SIN2W} & $0.23110323$ & \textsc{Calculated} \\
 & {\tt WMASS} & $79.9595 \ \mathrm{GeV}$ & \textsc{Calculated}\\
 & {\tt ZMASS} & $91.1876 \ \mathrm{GeV}$ & \textsc{Input}\\
&&&\\
\hline
&&&\\
 & {\tt FERMI\_CONST} & $1.16637 \times 10^{-5} \ \mathrm{GeV}^{-2}$ & \textsc{Input}\\
 & {\tt ALFA} & $7.7602239787 \times 10^{-3}$ & \textsc{Calculated}\\
\bf 2 & {\tt SIN2W} & $0.23119$ & \textsc{Input}\\
 & {\tt WMASS} & $ 79.9544 \ \mathrm{GeV}$ & \textsc{Calculated}\\
 & {\tt ZMASS} & $91.1876 \ \mathrm{GeV}$ & \textsc{Input}\\
&&&\\
\hline
&&&\\
 &{\tt FERMI\_CONST} & $1.16637 \times 10^{-5} \ \mathrm{GeV}^{-2}$ & \textsc{Input}\\
 &{\tt ALFA} & $7.5562544251\times 10^{-3}$ & \textsc{Calculated}\\
\bf 3 &{\tt SIN2W} & $0.22264585$ & \textsc{Calculated}\\
 &{\tt WMASS} & $80.3980 \ \mathrm{GeV}$ & \textsc{Input}\\
 & {\tt ZMASS} & $91.1876 \ \mathrm{GeV}$ & \textsc{Input}\\
&&&\\
\hline
&&&\\
 &{\tt  FERMI\_CONST} & $1.16637 \times 10^{-5} \ \mathrm{GeV}^{-2}$ & \textsc{Input}\\
 &{\tt  ALFA} & $7.2973525376 \times 10^{-3}$ & \textsc{Input}\\
\bf 4 &{\tt  SIN2W} & $0.23119$ & \textsc{Input}\\
 &{\tt  WMASS} & $80.398 \ \mathrm{GeV}$ & \textsc{Input}\\
 &{\tt  ZMASS} & $91.1876 \ \mathrm{GeV}$ & \textsc{Input}\\
&&&\\
\hline
\end{tabular}
\caption {\em  Electroweak input parameter schemes.}
\vspace{0.2cm}
\label{tab:Schemes}
\end{center}
\end{table}
\item {\tt FERMI\_CONST:} Fermi constant, used as input for the
  calculation of electroweak parameters. Default value is $1.16637
  \times 10^{-5} \ \mathrm{GeV}^{-2}$.
\item {\tt ALFA:} Fine structure constant, used as input for
  {\tt EWSCHEME} $=$ 1 and 4.  Within the other schemes this parameter is calculated. Default
  value is $7.2973525376 \times 10^{-3}$.
\item {\tt SIN2W:} Sinus squared of the weak mixing angle. Used as input for
  {\tt EWSCHEME} $= 2$ and 4. Within the other schemes this parameter is
  calculated. Default value is 0.23119.
\item {\tt WMASS:} $W$ boson mass in GeV. Default
  value is 80.398 GeV.
\item {\tt ZMASS:} $Z$ boson mass in GeV. Default
  value is 91.1876 GeV.
\item {\tt ANOM\_CPL:} Options for anomalous Higgs and gauge boson couplings. 
These are available for the  
  $Hjj$ and $W^+W^-jj$ production processes in VBF. 
  Default is set to {\tt .false.}.
\item {\tt KK\_MOD:} Option for the Warped Higgsless Model. 
 It is available for
  all $VVjj$ production modes in VBF. Default 
  is set to  {\tt .false.}.
\end{itemize}
\subsection{{\tt vbfnlo.dat} $-$ parameters for event output}
\label{sec:lha}
\textsc{Vbfnlo} generates parton level events
according to the most recent Les Houches Accord (LHA) format 
\cite{Alwall:2006yp} for processes available at leading order.
\begin{itemize}
\item {\tt LHA\_SWITCH:} Switch on or off output of LHA event files. 
Default is set to {\tt .false.}.
\item {\tt UNWEIGHTING\_SWITCH:} Option for event weights. If set to {\tt .true.}, 
  events are unweighted
  (event weight $=$ +1). If set to {\tt .false.}, events are weighted.  Default is set 
  to {\tt .false.}.
\item {\tt PRENEVUNW:} The number of events used in the last iteration in
  order to calculate/estimate  the premaximal weight which is needed in the
  first step of the unweighting procedure.   Default is 1000.
  After all events are unweighted, the maximal weight is again calculated and
  a reweighting procedure is applied.
\item {\tt TAUMASS:} Option to include the mass of $\tau$ leptons
 in the LHA file. So far this option only works for the {\tt vbf} processes. 
Default is set to {\tt .false.}.
\end{itemize}
\subsection{{\tt vbfnlo.dat} $-$ PDF  parameters}
\label{sec:pdf}
\textsc{Vbfnlo} may use built-in parton distribution functions (PDF)
or the LHAPDF library.  
\begin{itemize}
\item {\tt PDF\_SWITCH:} Option to choose which PDFs are used.
If set to {\tt 0}, built-in 
PDFs (CTEQ6L1 for LO and CTEQ6M for NLO calculations) are
used \cite{Pumplin:2002vw}. If set to {\tt 1},  an interface to LHAPDF is 
provided via LHAGLUE \cite{Whalley:2005nh}.
\item {\tt LO\_PDFSET:} LHAGLUE number for the LO PDF set, see PDFsets.index or
  Ref.~\cite{Whalley:2005nh}. Default is 10042 (CTEQ6L1). 
\item {\tt NLO\_PDFSET:}  LHAGLUE number for the NLO PDF set, see PDFsets.index or
  Ref.~\cite{Whalley:2005nh}. Default is 10000 (CTEQ6M). 
\end{itemize}


\subsection{{\tt vbfnlo.dat} $-$  parameters for histograms}
\label{sec:hist}

\textsc{Vbfnlo} provides output for histograms in the following formats:
\textsc{Topdrawer}\footnote{\tt http://www.pa.msu.edu/reference/topdrawer-docs/},
\textsc{Root}\footnote{\tt http://root.cern.ch/} and
\textsc{Gnuplot}\footnote{\tt http://www.gnuplot.info/}. 
\begin{itemize}
\item {\tt ROOT:} Enable output of histograms in \textsc{Root}
  format. Default is set to {\tt .false.}.
\item {\tt TOP:} Enable output of histograms  in 
  \textsc{Topdrawer} format. Default is set to {\tt .false.}.
\item {\tt GNU:} Enable output of histograms in \textsc{Gnuplot}
  format. Default is set to {\tt .true.}.
\item {\tt REPLACE:} Switch to overwrite existing histogram output files. 
Default is set to {\tt .true.}. 
\item {\tt ROOTFILE:} Name of the \textsc{Root} output file. Default is 
{\tt histograms}.
\item {\tt TOPFILE:}  Name of the \textsc{Topdrawer} output file. Default is 
{\tt histograms}.
\item {\tt GNUFILE:}  Name of the \textsc{Gnuplot} output file. Default is 
{\tt histograms}.
\end{itemize}


\subsection{{\tt cuts.dat} $-$ parameters for kinematical cuts}
\label{sec:cuts}
Jet-specific cuts:
\begin{itemize}
\item {\tt RJJ\_MIN:} Minimum separation of two identified jets, $\Delta
  R_{jj}= \sqrt{\Delta y^{2}_{jj}+\Delta \phi^{2}_{jj}}$, 
 used by the $k_\perp$ jet finding
  algorithm~\cite{Seymour:1997kj} which combines all
  partons.  Default is 0.8.
\item {\tt Y\_P\_MAX:} Maximum allowed pseudorapidity for final state 
partons.  Default is 5.0.
\item {\tt PT\_JET\_MIN:} Minimum transverse momentum
for identified jets. Default is 20 GeV.
\item {\tt Y\_JET\_MAX:} Maximum allowed rapidity for identified jets. 
Default is 4.5.
\end{itemize}
\vspace{0.1cm}
Lepton specific cuts:
\begin{itemize}
\item {\tt Y\_L\_MAX:} Maximum pseudorapidity for charged leptons. Default is 2.5.
\item {\tt PT\_L\_MIN:} Minimum transverse momentum for charged leptons. Default
  is 10 GeV.
\item {\tt MLL\_MIN:} Minimum invariant mass for any combination of oppositely 
  charged leptons. Default is 15 GeV.
\item {\tt RLL\_MIN:} Minimum separation of charged lepton pairs, $\Delta R_{\ell\ell}$. 
Default is 0.
\item {\tt RLL\_MAX:} Maximum separation of  charged lepton pairs, $\Delta R_{\ell\ell}$. 
Default is 50.
\end{itemize}
Photon specific cuts: 
\begin{itemize} 
\item {\tt Y\_G\_MAX:} Maximum pseudorapidity for photons. Default is 1.5.
\item {\tt PT\_G\_MIN:} Minimum transverse momentum for photons. Default
  is 20 GeV.
\item {\tt RGG\_MIN:} Minimum separation of photon pairs, $\Delta R_{\gamma\gamma}$. 
Default is 0.
\item {\tt RGG\_MAX:} Maximum separation of photon pairs, $\Delta R_{\gamma\gamma}$. 
Default is 50.
\end{itemize}
Additional cuts:
\begin{itemize}
\item {\tt RJL\_MIN:} Minimum separation of an identified jet and a charged lepton, $\Delta
  R_{j\ell}$. Default is 0. 
\item {\tt RJG\_MIN:}  Minimum separation of an identified jet and a photon, $\Delta R_{j\gamma}$. Default is~0. 
\item {\tt RLG\_MIN:}  Minimum separation of a charged lepton and a photon, $\Delta R_{\ell\gamma}$. Default is~0. 
\end{itemize}
VBF specific cuts\footnote{These apply only to {\tt vbf} and {\tt ggf} processes.}:
\begin{itemize}
\item {\tt ETAJJ\_MIN:} Minimum required pseudorapidity gap, $\Delta
  \eta_{jj}$, between  two tagging jets (the two leading jets in a $\pt$ ordering). Default  is 0.
\item {\tt YSIGN:} If set to {\tt .true.}, the two tagging jets are required
  to be found in the  opposite detector hemispheres. Default is {\tt
  .false.}.
\item {\tt LRAPIDGAP:} If set to {\tt .true.} all charged leptons are
required to lie between the two tagging jets in rapidity. Default is {\tt .false.}.
\item {\tt DELY\_JL:} Minimum rapidity distance of the charged leptons from the tagging jets,
  if {\tt LRAPIDGAP} is set to {\tt .true.}.  Default is 0.
\item {\tt GRAPIDGAP:} If set to {\tt .true.} all photons are
required to lie between the two tagging jets in rapidity. Default is {\tt .false.}.
\item {\tt DELY\_JG:} Minimum rapidity distance of photons from tagging jets,
  if {\tt GRAPIDGAP} is set to {\tt .true.}.  Default is 0.
\item {\tt MDIJ\_MIN:} Minimum dijet invariant mass of two tagging
  jets. Default is 0 GeV.
\item {\tt MDIJ\_MAX:} Maximum dijet invariant mass of
two tagging jets. Default is 14000~GeV. 
\item {\tt JVETO:} If set to {\tt .true.}, a central jet veto is applied. Default is
{\tt .false.}.
\item {\tt DELY\_JVETO:} Minimum rapidity separation of a central jet from two
tagging jets. Default is 0.
\item {\tt YMAX\_VETO:} Maximum pseudorapidity of a central jet. Default is 4.5.
\item {\tt PTMIN\_VETO:} Minimum transverse momentum of a central jet. Default is 10 GeV.
\end{itemize}


\subsection{Parameters for anomalous couplings}

\textsc{Vbfnlo} supports anomalous $HVV$ couplings, where $V=W,Z,\gamma$, in
both the production and the decay of a Higgs boson in VBF type reactions,  
i.e. for  \textsc{ProcId}s 100-107.
The anomalous $HVV$ couplings can be parameterized in the {\tt anom\_HVV.dat} 
input file. Moreover, the triple and quartic anomalous gauge boson  couplings 
for the VBF process $pp \to W^+ W^- jj(j)$ are included \cite{anomWW}.
These can be set in the {\tt anom\_WW.dat} file. 


\subsubsection{{\tt anom\_HVV.dat} $-$ anomalous $HVV$ couplings}
\label{sec:HVV}
Among the anomalous coupling input parameters, the user can choose
between three different parameterizations.
\begin{enumerate}
\item  A parameterization in terms of couplings in the effective Lagrangian 
  approach. In \textsc{Vbfnlo} effective dimension five operators are included
  corresponding to 
\begin{equation}
\label{eq:efflag}
{\cal L}^{d=5}_{\rm eff}= \frac{g_{5e}^{HVV}}{\Lambda_{5e}} H V^+_{\mu\nu}V^{-
  \mu\nu} + \frac{g_{5o}^{HVV}}{\Lambda_{5o}} H \tilde{V}^+_{\mu\nu} V^{- \mu\nu} \ ,
\end{equation}
where the subscript {\em e} or {\em o} refers to the $\cal{CP}$-even
or $\cal{CP}$-odd nature of the individual operators \cite{Figy:2004pt}.
\begin{itemize} 
\item {\tt PARAMETR1:} Parameter which switches on the effective Lagrangian
  parameterization Eq.~(\ref{eq:efflag}). The default value is {\tt .false.}.
\item {\tt LAMBDA5:} Mass scales $\Lambda_{5e}$ and
  $\Lambda_{5o}$ in units of GeV with 480 GeV chosen as default.
\item {\tt G5E\_HWW, G5E\_HZZ, G5E\_HGG, G5E\_HGZ:} Parameters which determine the couplings
  $g_{5e}^{HVV}$ of the $\cal{CP}$-even dimension five operators. Their default 
  values are set to  0.
\item {\tt G5O\_HWW, G5O\_HZZ, G5O\_HGG, G5O\_HGZ:} Parameters which determine the couplings
  $g_{5o}^{HVV}$ of the $\cal{CP}$-odd dimension five operators. Their default 
  values are set to 0.
\end{itemize}
\item The parameterization of the anomalous couplings by the L3 Collaboration
as given in Ref.~\cite{Achard:2004kn}. The parameters are $d$, $d_B$, $\Delta
g_1^Z$ and $\Delta \kappa_\gamma$. For the $\cal{CP}$-odd operators only three
 parameters are
needed. These are defined in analogy to the $\cal{CP}$-even ones and can be
related to the coefficients $f_i$ of the operators $\mathcal{O}_i$ 
in the effective Lagrangian as described in Ref. \cite{Hankele:2006ma} in the following way:
\begin{align} \label{eq:param2}
\begin{split}
&d = -\frac{m_W^2}{\Lambda^2}\ f_{WW}\,, \hspace{3.9cm} \tilde{d} =
-\frac{m_W^2}{\Lambda^2}\ f_{\tilde{W}W}\,,\\
&d_B = - \frac{m_W^2}{\Lambda^2}\
\frac{\sin^2{\theta_w}}{\cos^2{\theta_w}}\ f_{BB}\,, \hspace{2.3cm}
\tilde{d}_B = -\frac{m_W^2}{\Lambda^2}\
\frac{\sin^2{\theta_w}}{\cos^2{\theta_w}}\ f_{\tilde{B}B}\,,\\
&\Delta \kappa_\gamma = \kappa_\gamma -1 = \frac{m_W^2}{2 \Lambda^2} \
(f_B + f_W)\,, \hspace{1cm} \tilde{\kappa}_\gamma = \frac{m_W^2}{2
  \Lambda^2}\ f_{\tilde{B}}\,,\\
&\Delta g_1^Z = g_1^Z -1 = \frac{m_Z^2}{\Lambda^2}\, \ \frac{f_W}{2}\,
\end{split}
\end{align}
\begin{itemize}
\item {\tt PARAMETR2:} Parameter which switches on the above mentioned parameterization.
 The default  value is  {\tt .false.}.
\item {\tt D\_EVEN, DB\_EVEN, DG1Z\_EVEN, DKGAM\_EVEN:} Parameters which are 
  the $\cal{CP}$-even couplings in this parameterization with default 0.
\item {\tt D\_ODD, DB\_ODD, KGAM\_ODD:} Parameters which are  
the $\cal{CP}$-odd couplings in this parameterization with default values
equal to 0.
\item  {\tt HVV1:} Parameter which determines which anomalous $HVV$
  couplings are used for the run. For {\tt HVV1} $=$ 0, only the $HZ\gamma$ couplings,
  for {\tt HVV1} $=$ 1, only the $H\gamma\gamma$ coupling, for {\tt HVV1}
  $=$ 2, only the $HZZ$ coupling and for {\tt HVV1} $=$ 3, only the
  $HWW$ coupling is used. If {\tt HVV1} is set to 4, all possible
  anomalous couplings are used. This is also the default value.
\end{itemize}
\item  The parameterization of the anomalous couplings in terms of coefficients 
$f_i/\Lambda^2$ of the operators ${\cal O}_{WW}$,  ${\cal O}_{BB}$, ${\cal O}_W$ 
and ${\cal O}_B$ and their corresponding $\cal{CP}$-odd 
operators according to Refs.~\cite{Hagiwara:1993qt,Hagiwara:1993ck}.
\begin{itemize}
\item {\tt PARAMETR3:} Parameter which switches on the parameterization stated above.
  The default value is {\tt .false.}.
\item {\tt FWW\_EVEN, FBB\_EVEN, FW\_EVEN, FB\_EVEN:} Parameters which 
 represent the coefficients 
  of the $\cal{CP}$-even operators with default values equal to 0.
\item {\tt FWW\_ODD, FBB\_ODD, FB\_ODD:} Parameters which are the coefficients
  of the $\cal{CP}$-odd
  operators with default values 0.
\item {\tt HVV2:} Parameter which allows to choose which anomalous HVV
  couplings are used. For {\tt HVV2} = 0, only the $HZ\gamma$ coupling,
  for {\tt HVV2} = 1, only the $H\gamma\gamma$ coupling, for {\tt HVV2}
  = 2, only the $HZZ$ coupling and for {\tt HVV2} = 3, only the
  $HWW$ coupling is used. If set to 4 all possible
  anomalous couplings are used. The default value is 4.
\item {\tt TREEFAC:} Parameter that multiplies the $HVV$
  tensor present in the SM Lagrangian. Default is 1.
\item {\tt LOOPFAC:} Parameter that multiplies the $HZ\gamma$ and $H\gamma\gamma$
  vertices induced by SM loops. The default is chosen to be 1.
\end{itemize}
\end{enumerate}
Moreover, for all parameterizations two different form factors can be chosen
as described in Refs.~\cite{Figy:2004pt,Hankele:2006ma}. They model effective, 
momentum dependent $HVV$ vertices, motivated from new physics
entering with a large scale $\Lambda$ at the loop level.
\begin{eqnarray}
\label{eq:ff1}
F_1 &=& \frac{\Lambda^2}{q_1^2 - \Lambda^2}\ \frac{\Lambda^2}{q_2^2 -
  \Lambda^2}\,, 
\\
\label{eq:ff2}
F_2 &=& -2 \,\Lambda^2 \, C_0\!\left(q_1^2, q_2^2, (q_1+q_2)^2,
\Lambda^2\right). 
\end{eqnarray}
Here the $q_i$ are the momenta of the vector bosons and 
$C_{0}$ is the scalar one-loop three point function 
in the notation of Ref.\cite{Passarino:1978jh}.
\begin{itemize}
\item {\tt FORMFACTOR:} Parameter which switches on the above 
 form factor parameterization. The default value is set to {\tt .false.}.
\item {\tt MASS\_SCALE:} Characteristic mass scale of new physics 
 $\Lambda$ in units of GeV. The default value is 200 GeV.
\item {\tt FFAC:} Parameter which is used to select one particular form factor 
 out of Eqs.~(\ref{eq:ff1}) and~(\ref{eq:ff2}). If {\tt FFAC} $=$ 1,
  the form factor $F_{1}$ is used for the parameterization.
  {\tt FFAC} $=$ 2 selects $F_{2}$, which is also the default value.
\end{itemize}
Finnaly, the two parameters can be used to rescale the SM $HVV$ couplings.
\begin{itemize}
\item {\tt TREEFAC:} Parameter that multiplies the $HVV$
  tensor present in the SM Lagrangian. Default is 1.
\item {\tt LOOPFAC:} Parameter that multiplies the $HZ\gamma$ and $H\gamma\gamma$
  vertices induced by SM loops. The default is chosen to be 1.
\end{itemize}

\subsubsection{{\tt anom\_WW.dat} $-$ anomalous triple and quartic  gauge
  boson couplings}
\label{sec:TGC}
The triple and quartic anomalous gauge boson couplings can be set in 
{\tt anom\_WW.dat}. The input
values are the coefficients $f_i/\Lambda^2$ of the $\cal{CP}$-even operators in 
the effective Lagrangian
${\cal O}_{BW}$,  ${\cal O}_{DW}$, ${\cal O}_{WWW}$, ${\cal O}_{WW}$,  ${\cal
  O}_{BB}$, ${\cal O}_W$ and ${\cal O}_B$ and their corresponding $\cal{CP}$-odd
operators as described in Refs.~\cite{Buchmuller:1985jz, Hagiwara:1993ck}.
\begin{itemize}
\item {\tt FBW, FDW, FWWW, FWW, FBB, FW, FB:} Parameters which give the values of the 
 coefficients of the $\cal{CP}$-even operators. The default values for these parameters are~0.
\item {\tt FWWt, FBWt, FBBt, FWt, FBt, FWWWt, FDWt:} Parameters which are the coefficients 
of the $\cal{CP}$-odd operators. The default values for these parameters are 0.
\item {\tt OVS:} Parameter which switches on the overall factor scheme. The
  default value is {\tt .false.}.
\item {\tt FORMFAC:} Parameter which allows to include the form factor
\begin{equation}
\label{eq:ff3}
F= \(1 + \frac{s}{\Lambda^2}\)^{-n} \, ,
\end{equation}
in the calculation, where $\Lambda$ is a characteristic new physics mass scale.
The default value is {\tt .false.}.
\item {\tt LAMBDA:} Parameter that gives the above scale $\Lambda$ in units of GeV. 
Default is 2500~GeV.
\item {\tt EXPFAC:} Parameter to set the exponent $n$ in Eq.~(\ref{eq:ff3})
 with default value 2.
\end{itemize}


\subsection{{\tt kk\_input.dat} $-$ parameters for the Warped Higgsless model}

\textsc{Vbfnlo} is capable of calculating the weak boson fusion processes
$VV$+2~jets in the Warped Higgsless scenario \cite{Csaki:2003zu} at LO and
NLO QCD level (see, for example, Ref.~\cite{Englert:2008tn} for a
phenomenological application).  The model parameters can be generated by
\textsc{Vbfnlo} via the input file {\tt kk\_input.dat} for a hard wired
choice of the relevant five dimensional gauge parameters. The input values
are
\begin{itemize}
\item Switch that determines whether \textsc{Vbfnlo} should calculate
the  parameters needed by the model. Default is {\tt .true.}.
\item  Location of the UV brane for the generation of the model
parameters in the Warped Higgsless Model. Default is $R=9.75\times 10^{-9}$, 
which amounts to the Kaluza-Klein excitations having masses of $m_{W_2}=700~\mathrm{GeV}$, 
$m_{Z_2}=695~\mathrm{GeV}$, and $m_{Z'_1}=718~\mathrm{GeV}$. Smaller values of
$R$ result in a heavier Kaluza-Klein spectrum.
\item The maximum number of Kaluza-Klein $W^\pm_k$ states to be included on 
top of the Standard Model $W^\pm$ bosons, which correspond to $W^\pm_{k=1}$.    
All states $k\geq 3$ are phenomenologically irrelevant. 
Default is 1.
\item The maximum number of Kaluza-Klein $Z_k$ states to be included on top of 
the Standard Model $Z$ boson, which corresponds to $Z_{k=1}$.
All states $k\geq 3$ are phenomenologically irrelevant. 
Default is 1.
\item The maximum number of Kaluza-Klein $Z'_k$ bosons that are excitations of 
the Standard Model photon $Z'_{k=0}$. States $k\geq 2$ are phenomenologically
irrelevant. Default is 1.
\end{itemize}
The explicit breaking of higher dimensional gauge invariance is balanced 
according to the description of \cite{KK,Englert:2008wp}, 
where also more details on the model and its implementation can be found.

\textsc{Vbfnlo} generates the text
file {\tt kk\_coupl\_inp.dat}, which documents the calculated model parameters, 
i.e.\ Kaluza-Klein gauge boson masses, couplings and widths of the specified
input parameters. This file can also be used as 
input file for advanced users who want to run the code with 
their own set of parameters. To that end, select {\tt .false.} in the file 
{\tt kk\_input.dat}. \textsc{Vbfnlo} will then calculate the gauge boson
widths on the  basis of these parameters by the decay to the lower lying states.
Information on the widths and on the sum rules relating the various
gauge boson couplings \cite{Csaki:2003dt} are written to {\tt kk\_check.dat}.


\subsection{{\tt ggflo.dat} $-$ general parameters for gluon fusion processes}

In \textsc{Vbfnlo}, the double real-emission corrections to $gg\to \phi$,  which lead to
$\phi$+ 2~jet events at order $\alpha_s^{4}$, are  included. Here, $\phi$ can
be a scalar ($h,H$) or pseudo-scalar ($A$) Higgs boson as in a generic two-Higgs
-doublet model (2HDM) of type II.  Contributions contain top- and
bottom-quark triangles, boxes and pentagon diagrams, i.e. the full mass
dependence of the loop induced production. Interference effects between loops 
with bottom
and top quarks as well as between ${\cal{CP}}$-even and ${\cal{CP}}$-odd
couplings of the heavy quarks are fully taken into account. An option to use
the large top mass approximation, which works well for intermediate
Higgs boson masses, provided that the transverse momenta of the final state
partons are smaller than the top quark mass, is also implemented.
  
Higgs boson plus two jets production via gluon fusion requires usage of the
{\tt ggflo} executable program rather than the {\tt vbfnlo} one. Moreover, 
the {\tt vbfnlo.dat} input file does not need to be modified in order to set
the correct value of process ID. In order to be more
transparent to the user, this information is automatically assumed when
running {\tt ggflo}. However, the following additional parameters have to be
adjusted in the {\tt ggflo.dat} file:
\begin{table}[t!]
\newcommand{\lstrut}{{$\strut\atop\strut$}}
\begin{center}
\begin{tabular}{c|p{14cm}}
\hline
&\\
\textsc{ID} & \textsc{Model Description} \\
&\\
\hline
&\\
\bf 0 & Toy model with settings provided by the user \\
\bf 1 & $\cal{CP}$-even Higgs ($H_{\rm SM}$) in the large top quark mass limit 
($m_{t} \to \infty$)\\
\bf 2 & $\cal{CP}$-odd Higgs ($H_{\rm SM}$) in the large top quark mass limit 
($m_{t} \to \infty$) \\
\bf 3 & $\cal{CP}$-even Higgs ($H_{\rm SM}$)  with full mass dependence of the 
top quark loop \\ 
\bf 4 & $\cal{CP}$-even Higgs ($H_{\rm SM}$)  with full mass dependence of the 
bottom quark loop \\ 
\bf 5 & $\cal{CP}$-even Higgs ($H_{\rm SM}$)  with full mass dependence of the 
top and bottom quark loop\\ 
\bf 6 & $\cal{CP}$-odd Higgs (A) with full mass dependence of 
the top quark loop\\ 
\bf 7 & $\cal{CP}$-odd Higgs (A) with full mass dependence of 
the bottom quark loop \\ 
\bf 8 & $\cal{CP}$-odd Higgs (A) with full mass dependence of 
the top and
bottom quark loop\\ 
\bf 9 & $\cal{CP}$-even  Higgs (h) with full mass dependence of 
the top quark loop\\ 
\bf 10 & $\cal{CP}$-even Higgs (h) with full mass dependence of 
the bottom quark loop \\ 
\bf 11 & $\cal{CP}$-even Higgs (h) with full mass dependence of 
the top and
bottom quark loop\\ 
\bf 12 & $\cal{CP}$-even Higgs (H) with full mass dependence of 
the top quark loop\\ 
\bf 13 & $\cal{CP}$-even Higgs (H) with full mass dependence of 
the bottom quark loop \\ 
\bf 14 & $\cal{CP}$-even Higgs (H) with full mass dependence of 
the top and bottom quark loop\\ 
&\\
\hline
\end{tabular}
\caption {\it Different models for the production of a Higgs boson plus
 two jets via gluon fusion.}
\vspace{0.2cm}
\label{ggf:models}
\end{center}
\end{table}
\begin{itemize}
\item {\tt PROCESSGGF:} Model ID. A summary is given in 
Table~\ref{ggf:models}. 
Default is 3.
\item {\tt SUBPRQQ:} Switch for the subprocesses with quark-quark initial
  state. Default is set to {\tt .true.}.  
\item {\tt SUBPRQG:} Switch for the subprocesses with quark-gluon initial
  state. Default is set to {\tt .true.}.  
\item {\tt SUBPRGG:} Switch for the subprocesses with gluon-gluon initial
  state. Default is set to {\tt .true.}.  
\item {\tt TAN\_BETA:} Ratio of vacuum expectation values, $\tan\beta =
  v_u/v_d$ where $v^2 = v_u^2 + v_d^2$.
\item {\tt ALPHA:} Neutral Higgs boson mixing angle $\alpha$, which arises when
  the $\cal{CP}$-even Higgs boson mass matrix is diagonalized to obtain the physical 
  $\cal{CP}$-even Higgs boson states, $h$ and $H$.
\end{itemize}


\section{Checks}

Extensive checks for the LO and the real emission amplitudes as well as for
the total LO cross sections have been performed for all processes implemented in \textsc{Vbfnlo}.  Born amplitudes and 
real emission diagrams have been compared  with the fully automatically
generated results provided by  \textsc{MadGraph}\cite{Stelzer:1994ta}.
Complete agreement has been found in each case.  Moreover, total LO cross
sections with a minimal set of cuts agree with the respective results obtained
by \textsc{MadEvent}\footnote{\tt http://madgraph.hep.uiuc.edu/}
\cite{Maltoni:2002qb,Alwall:2007st} and
\textsc{Helac-Phegas}\footnote{\tt http://helac-phegas.web.cern.ch/helac-phegas/}
\cite{Kanaki:2000ey,Papadopoulos:2005ky,Cafarella:2007pc}, a completely  automatic parton level
event generator  based on Dyson-Schwinger recursive equations. 

All LHA event
files for the LO processes have been tested with \textsc{Herwig++}\footnote{\tt
http://projects.hepforge.org/herwig/} \cite{Bahr:2008pv},  a general purpose
Monte Carlo event generator for the simulation of hard lepton-lepton and
hadron-hadron collisions.

As a final and very important test, comparisons with already published
results have been made. In Ref.~\cite{:2008uu}, a tuned comparison of LO and
NLO QCD results for Higgs boson production via vector boson fusion at the LHC
has been performed. Three different calculations have been cross checked:
\textsc{Vbfnlo}, the results of
Refs.~\cite{Ciccolini:2007jr,Ciccolini:2007ec}, and the \textsc{VV2H}
program\footnote{\tt http://people.web.psi.ch/spira/vv2h/}. For the dominant
$t$- and $u$-channel contributions which are implemented in \textsc{Vbfnlo},
good agreement has been found.  For the triboson processes a comparison for
the production of on-shell gauge bosons without leptonic decays has been
performed with the results presented in Ref.~\cite{Binoth:2008kt}. Again,
good agreement has been found.
Results for the $\cal{CP}$-odd and $\cal{CP}$-even Higgs boson production
via gluon fusion have been tested against \textsc{FeynArts}\footnote{\tt
  http://www.feynarts.de/} \cite{Hahn:2000kx,Hahn:2001rv}.

\section{Summary \& Outlook}

\textsc{Vbfnlo} is a fully flexible partonic Monte Carlo program for vector
boson fusion, double and triple vector boson production processes at NLO QCD
accuracy.  The simulation of $\cal{CP}$-even and $\cal{CP}$-odd Higgs boson
production in gluon fusion, associated with two additional jets, is
implemented at leading order (for this process only, the LO starts at
one-loop level).

Future improvements are directed along two main lines of development:
Further processes at NLO QCD accuracy will be included (e.g., $pp\to WW\gamma$
and $pp\to W\gamma j$) and new features will be added to the already existing
processes, such as anomalous triple and quartic gauge boson couplings and 
Kaluza-Klein excitations. 
Higgs production via gluon fusion within the generic two-Higgs-doublet model 
will be extended to a complete simulation within the Minimal Supersymmetric
Standard Model (MSSM), including full dependence on scalar top and
bottom quarks masses.
Matching the NLO QCD processes to a parton shower at next-to-leading
logarithmic accuracy is currently in progress.


\section*{Acknowledgments}

The research presented in this article was supported in part by the Deutsche Forschungsgemeinschaft
via the Sonderforschungsbereich/Transregio  SFB/TR-9 ``Computational Particle
Physics'' and the Graduiertenkolleg ``High Energy Physics and Particle
Astrophysics'', and in part by the National Science Foundation under
Grant No.~PHY05-51164. 

F.~Campanario acknowledges a postdoctoral fellowship of the
Generalitat Valenciana (Beca Postdoctoral d`Excell\'encia) and S.~Pl\"atzer support from the Landesgraduiertenf\"orderung
Baden-W\"urttemberg.
M.~Worek was funded in part by the RTN European Programme 
MRTN-CT-2006-035505 HEPTOOLS - Tools and Precision Calculations 
for Physics Discoveries at Colliders and B.~J\"ager by the Initiative and Networking Fund of the
Helmholtz Association, contract HA-101 ("Physics at the Terascale"). 


\begingroup\begin{thebibliography}{10}

\bibitem{CO}
C.~Georg, ``Interferenzeffekte in Vektorboson-Fusion'', {Diploma Thesis, ITP
  Karlsruhe 2005}.

\bibitem{Andersen:2007mp}
J.~R. Andersen, T.~Binoth, G.~Heinrich, and J.~M. Smillie, ``{Loop induced
  interference effects in Higgs Boson plus two jet production at the LHC}'',
  {\em JHEP} {\bf 0802} (2008) 057,
\href{http://www.arXiv.org/abs/0709.3513}{{\tt arXiv:0709.3513 [hep-ph]}}.

\bibitem{Bredenstein:2008tm}
A.~Bredenstein, K.~Hagiwara, and B.~Jager, ``{Mixed QCD-electroweak
  contributions to Higgs-plus-dijet production at the LHC}'', {\em Phys. Rev.}
  {\bf D77} (2008) 073004,
\href{http://www.arXiv.org/abs/0801.4231}{{\tt arXiv:0801.4231 [hep-ph]}}.

\bibitem{Figy:2003nv}
T.~Figy, C.~Oleari, and D.~Zeppenfeld, ``Next-to-leading order jet
  distributions for Higgs boson production via weak boson fusion'', {\em Phys.
  Rev.} {\bf D68} (2003) 073005,
\href{http://www.arXiv.org/abs/hep-ph/0306109}{{\tt hep-ph/0306109}}.

\bibitem{Figy:2007kv}
T.~Figy, V.~Hankele, and D.~Zeppenfeld, ``{Next-to-leading order QCD
  corrections to Higgs plus three jet production in vector-boson fusion}'',
  {\em JHEP} {\bf 0802} (2008) 076,
\href{http://www.arXiv.org/abs/0710.5621}{{\tt arXiv:0710.5621 [hep-ph]}}.

\bibitem{Oleari:2003tc}
C.~Oleari and D.~Zeppenfeld, ``QCD corrections to electroweak $l\nu_ljj$
and $l^+l^-jj$ production'', {\em Phys. Rev.} {\bf D69} (2004)
  093004,
\href{http://www.arXiv.org/abs/hep-ph/0310156}{{\tt hep-ph/0310156}}.

\bibitem{Jager:2006zc}
B.~Jager, C.~Oleari, and D.~Zeppenfeld, ``Next-to-leading order QCD corrections
  to $W^+ W^-$ production via vector-boson fusion'', {\em JHEP} {\bf 0607} 
(2006)  015,
\href{http://www.arXiv.org/abs/hep-ph/0603177}{{\tt hep-ph/0603177}}.

\bibitem{Jager:2006cp}
B.~Jager, C.~Oleari, and D.~Zeppenfeld, ``Next-to-leading order QCD corrections
  to Z boson pair production via vector-boson fusion'', {\em Phys. Rev.} {\bf
  D73} (2006) 113006,
\href{http://www.arXiv.org/abs/hep-ph/0604200}{{\tt hep-ph/0604200}}.

\bibitem{Bozzi:2007ur}
G.~Bozzi, B.~Jager, C.~Oleari, and D.~Zeppenfeld, ``Next-to-leading order QCD
  corrections to $W^+Z$ and $W^-Z$ production via vector-boson 
fusion'', {\em Phys.  Rev.} {\bf D75} (2007) 073004,
\href{http://www.arXiv.org/abs/hep-ph/0701105}{{\tt hep-ph/0701105}}.

\bibitem{Hankele:2007sb}
V.~Hankele and D.~Zeppenfeld, ``{QCD corrections to hadronic WWZ production
  with leptonic decays}'', {\em Phys. Lett.} {\bf B661} (2008) 103,
\href{http://www.arXiv.org/abs/0712.3544}{{\tt arXiv:0712.3544 [hep-ph]}}.

\bibitem{Campanario:2008yg}
F.~Campanario, V.~Hankele, C.~Oleari, S.~Prestel, and D.~Zeppenfeld, ``{QCD
  corrections to charged triple vector boson production with leptonic decay}'',
\href{http://www.arXiv.org/abs/0809.0790}{{\tt arXiv:0809.0790 [hep-ph]}}.

\bibitem{DelDuca:2001eu}
V.~Del~Duca, W.~Kilgore, C.~Oleari, C.~Schmidt, and D.~Zeppenfeld, ``{H + 2
  jets via gluon fusion}'', {\em Phys. Rev. Lett.} {\bf 87} (2001) 122001,
\href{http://www.arXiv.org/abs/hep-ph/0105129}{{\tt hep-ph/0105129}}.

\bibitem{DelDuca:2001fn}
V.~Del~Duca, W.~Kilgore, C.~Oleari, C.~Schmidt, and D.~Zeppenfeld,
  ``{Gluon-fusion contributions to H + 2 jet production}'', {\em Nucl. Phys.}
  {\bf B616} (2001) 367,
\href{http://www.arXiv.org/abs/hep-ph/0108030}{{\tt hep-ph/0108030}}.

\bibitem{DelDuca:2006hk}
V.~Del~Duca {\em et al.}, ``{Monte Carlo studies of the jet activity in Higgs +
  2 jet events}'', {\em JHEP} {\bf 0610} (2006) 016,
\href{http://www.arXiv.org/abs/hep-ph/0608158}{{\tt hep-ph/0608158}}.

\bibitem{Klamke:2007cu}
G.~Klamke and D.~Zeppenfeld, ``{Higgs plus two jet production via gluon fusion
  as a signal at the CERN LHC}'', {\em JHEP} {\bf 0704} (2007) 052,
\href{http://www.arXiv.org/abs/hep-ph/0703202}{{\tt hep-ph/0703202}}.

\bibitem{GF}
M.~Kubocz, ``Produktion des CP-ungeraden Higgs-Bosons im Prozess $pp\to
  A^{0}jjX$'', {Diploma Thesis, ITP Karlsruhe 2006}.

\bibitem{Murayama:1992gi}
H.~Murayama, I.~Watanabe, and K.~Hagiwara, ``Helas: HELicity amplitude
  subroutines for Feynman diagram evaluations'', {\tt KEK-91-11}.

\bibitem{Pumplin:2002vw}
J.~Pumplin {\em et al.}, ``New generation of parton distributions with
  uncertainties from global QCD analysis'', {\em JHEP} {\bf 0207} (2002) 012,
\href{http://www.arXiv.org/abs/hep-ph/0201195}{{\tt hep-ph/0201195}}.

\bibitem{Amsler:2008zz}
{Particle Data Group} Collaboration, C.~Amsler {\em et al.}, ``{Review of
  Particle Physics}'', {\em Phys. Lett.} {\bf B667} (2008) 1.

\bibitem{Alwall:2006yp}
J.~Alwall {\em et al.}, ``A Standard format for Les Houches event files'', {\em
  Comput. Phys. Commun.} {\bf 176} (2007) 300,
\href{http://www.arXiv.org/abs/hep-ph/0609017}{{\tt hep-ph/0609017}}.

\bibitem{Whalley:2005nh}
M.~R. Whalley, D.~Bourilkov, and R.~C. Group, ``The Les Houches accord PDFs
  (LHAPDF) and LHAGLUE'',
\href{http://www.arXiv.org/abs/hep-ph/0508110}{{\tt hep-ph/0508110}}.

\bibitem{Seymour:1997kj}
M.~H. Seymour, ``{Jet shapes in hadron collisions: Higher orders, resummation
  and hadronization}'', {\em Nucl. Phys.} {\bf B513} (1998) 269,
\href{http://www.arXiv.org/abs/hep-ph/9707338}{{\tt hep-ph/9707338}}.

\bibitem{anomWW}
N.~Greiner, ``Anomale kopplungen bei der W-Paar-Produktion in
  Vektor-Boson-Fusion'', {Diploma Thesis, ITP Karlsruhe 2006}.

\bibitem{Figy:2004pt}
T.~Figy and D.~Zeppenfeld, ``QCD corrections to jet correlations in weak boson
  fusion'', {\em Phys. Lett.} {\bf B591} (2004) 297,
\href{http://www.arXiv.org/abs/hep-ph/0403297}{{\tt hep-ph/0403297}}.

\bibitem{Achard:2004kn}
{L3} Collaboration, P.~Achard {\em et al.}, ``Search for anomalous couplings in
  the Higgs sector at LEP'', {\em Phys. Lett.} {\bf B589} (2004) 89,
\href{http://www.arXiv.org/abs/hep-ex/0403037}{{\tt hep-ex/0403037}}.

\bibitem{Hankele:2006ma}
V.~Hankele, G.~Klamke, D.~Zeppenfeld, and T.~Figy, ``Anomalous Higgs boson
  couplings in vector boson fusion at the CERN LHC'', {\em Phys. Rev.} {\bf
  D74} (2006) 095001,
\href{http://www.arXiv.org/abs/hep-ph/0609075}{{\tt hep-ph/0609075}}.

\bibitem{Hagiwara:1993qt}
K.~Hagiwara, R.~Szalapski, and D.~Zeppenfeld, ``Anomalous Higgs boson
  production and decay'', {\em Phys. Lett.} {\bf B318} (1993) 155,
\href{http://www.arXiv.org/abs/hep-ph/9308347}{{\tt hep-ph/9308347}}.

\bibitem{Hagiwara:1993ck}
K.~Hagiwara, S.~Ishihara, R.~Szalapski, and D.~Zeppenfeld, ``Low-energy effects
  of new interactions in the electroweak boson sector'', {\em Phys. Rev.} {\bf
  D48} (1993)
2182.

\bibitem{Passarino:1978jh}
  G.~Passarino and M.~J.~G.~Veltman,
  ``One Loop Corrections For $e^+e^-$ Annihilation into $\mu^+\mu^-$ in the Weinberg
  Model,''
  {\em Nucl. Phys.}  {\bf B160} (1979) 151.

\bibitem{Buchmuller:1985jz}
W.~Buchmuller and D.~Wyler, ``Effective Lagrangian Analysis of New Interactions
  and Flavor Conservation'', {\em Nucl. Phys.} {\bf B268} (1986)
621.

\bibitem{Csaki:2003zu}
C.~Csaki, C.~Grojean, L.~Pilo, and J.~Terning, ``Towards a realistic model of
  Higgsless electroweak symmetry breaking'', {\em Phys. Rev. Lett.} {\bf 92}
  (2004) 101802,
\href{http://www.arXiv.org/abs/hep-ph/0308038}{{\tt hep-ph/0308038}}.

\bibitem{Englert:2008tn}
C.~Englert, B.~Jager, M.~Worek, and D.~Zeppenfeld, ``{Observing Strongly
  Interacting Vector Boson Systems at the CERN Large Hadron Collider}'',
\href{http://www.arXiv.org/abs/0810.4861}{{\tt arXiv:0810.4861 [hep-ph]}}.

\bibitem{KK}
C.~Englert, ``Spin 1 Resonances in Vector Boson Fusion in Warped Higgsless
  Models'', {Diploma Thesis, ITP Karlsruhe 2007}.

\bibitem{Englert:2008wp}
  C.~Englert, B.~Jager and D.~Zeppenfeld, 
``QCD Corrections to Vector-Boson Fusion Processes in Warped Higgsless
  Models'', 
\href{http://www.arXiv.org/abs/0812.2564}{{\tt arXiv:0812.2564 [hep-ph]}}.

\bibitem{Csaki:2003dt}
C.~Csaki, C.~Grojean, H.~Murayama, L.~Pilo, and J.~Terning, ``{Gauge theories
  on an interval: Unitarity without a Higgs}'', {\em Phys. Rev.} {\bf D69}
  (2004) 055006,
\href{http://www.arXiv.org/abs/hep-ph/0305237}{{\tt hep-ph/0305237}}.

\bibitem{Stelzer:1994ta}
T.~Stelzer and W.~F. Long, ``{Automatic generation of tree level helicity
  amplitudes}'', {\em Comput. Phys. Commun.} {\bf 81} (1994) 357,
\href{http://www.arXiv.org/abs/hep-ph/9401258}{{\tt hep-ph/9401258}}.

\bibitem{Maltoni:2002qb}
F.~Maltoni and T.~Stelzer, ``{MadEvent: Automatic event generation with
  MadGraph}'', {\em JHEP} {\bf 0302} (2003) 027,
\href{http://www.arXiv.org/abs/hep-ph/0208156}{{\tt hep-ph/0208156}}.

\bibitem{Alwall:2007st}
J.~Alwall {\em et al.}, ``{MadGraph/MadEvent v4: The New Web Generation}'',
  {\em JHEP} {\bf 0709} (2007) 028,
\href{http://www.arXiv.org/abs/0706.2334}{{\tt arXiv:0706.2334 [hep-ph]}}.

\bibitem{Kanaki:2000ey}
A.~Kanaki and C.~G.~Papadopoulos, 
``HELAC: A package to compute electroweak helicity amplitudes'',
{\em Comput. Phys. Commun.} {\bf 132} (2000) 306,
\href{http://www.arXiv.org/abs/hep-ph/0002082}{{\tt hep-ph/0002082}}.

\bibitem{Papadopoulos:2005ky}
C.~G. Papadopoulos and M.~Worek, ``{Multi-parton cross sections at hadron
  colliders}'', {\em Eur. Phys. J.} {\bf C50} (2007) 843,
\href{http://www.arXiv.org/abs/hep-ph/0512150}{{\tt hep-ph/0512150}}.

\bibitem{Cafarella:2007pc}
A.~Cafarella, C.~G. Papadopoulos, and M.~Worek, ``{Helac-Phegas: A Generator
  for all parton level processes}'',
\href{http://www.arXiv.org/abs/0710.2427}{{\tt arXiv:0710.2427 [hep-ph]}}.


\bibitem{Bahr:2008pv}
M.~Bahr {\it et al.},``Herwig++ Physics and Manual'', {\em Eur. Phys. J.} 
{\bf C58} (2008) 639, 
\href{http://www.arXiv.org/abs/0803.0883}{{\tt arXiv:0803.0883 [hep-ph]}}.

\bibitem{:2008uu}
N.~E. Adam {\em et al.}, ``{Higgs Working Group Summary Report}'',
\href{http://www.arXiv.org/abs/0803.1154}{{\tt arXiv:0803.1154 [hep-ph]}}.

\bibitem{Ciccolini:2007jr}
M.~Ciccolini, A.~Denner, and S.~Dittmaier, ``{Strong and electroweak
  corrections to the production of Higgs + 2 jets via weak interactions at the
  LHC}'', {\em Phys. Rev. Lett.} {\bf 99} (2007) 161803,
\href{http://www.arXiv.org/abs/0707.0381}{{\tt arXiv:0707.0381 [hep-ph]}}.

\bibitem{Ciccolini:2007ec}
M.~Ciccolini, A.~Denner, and S.~Dittmaier, ``{Electroweak and QCD corrections
  to Higgs production via vector-boson fusion at the LHC}'', {\em Phys. Rev.}
  {\bf D77} (2008) 013002,
\href{http://www.arXiv.org/abs/0710.4749}{{\tt arXiv:0710.4749}}.

\bibitem{Binoth:2008kt}
T.~Binoth, G.~Ossola, C.~G. Papadopoulos, and R.~Pittau, ``{NLO QCD corrections
  to tri-boson production}'', {\em JHEP} {\bf 0806} (2008) 082,
\href{http://www.arXiv.org/abs/0804.0350}{{\tt arXiv:0804.0350 [hep-ph]}}.

\bibitem{Hahn:2000kx}
T.~Hahn, ``{Generating Feynman diagrams and amplitudes with FeynArts 3}'', {\em
  Comput. Phys. Commun.} {\bf 140} (2001) 418,
\href{http://www.arXiv.org/abs/hep-ph/0012260}{{\tt hep-ph/0012260}}.

\bibitem{Hahn:2001rv}
T.~Hahn and C.~Schappacher, ``{The implementation of the minimal supersymmetric
  standard model in FeynArts and FormCalc}'', {\em Comput. Phys. Commun.} {\bf
  143} (2002) 54,
\href{http://www.arXiv.org/abs/hep-ph/0105349}{{\tt hep-ph/0105349}}.

\end{thebibliography}\endgroup

\providecommand{\href}[2]{#2}

\end{document}